\documentstyle[12pt] {article}
\newcommand{\beq}{\begin{equation}}
\newcommand{\eeq}{\end{equation}}
\newcommand{\beqa}{\begin{eqnarray}}
\newcommand{\eeqa}{\end{eqnarray}}
\newcommand{\ba}{\begin{array}}
\newcommand{\ea}{\end{array}}

\begin{document}

\begin{center}
{\large \bf Accuracy of the Semi--Classical Approximation:\\
the Pullen Edmonds Hamiltonian}\footnote{This work has been partially
supported by the Ministero dell'Universit\`a e della Ricerca Scientifica
e Tecnologica (MURST).}

\vskip 0.5 truecm

{\bf S. Graffi}\\
Dipartimento di Matematica dell'Universit\`a di Bologna,\\
Piazza di Porta S. Donato 5, I 40127, Bologna, Italy

\vskip 0.5 truecm

{\bf V.R. Manfredi}\footnote{Author to whom all correspondence and reprint
requests should be addressed. E--Mail: VAXFPD::MANFREDI,
MANFREDI@PADOVA.INFN.IT}
\\
Dipartimento di Fisica "G. Galilei" dell'Universit\`a di Padova,\\
INFN, Sezione di Padova,\\
Via Marzolo 8, I 35131 Padova, Italy\footnote{Permanent address}
\\
Interdisciplinary Laboratory, International School for Advanced Studies,\\
Strada Costiera, 11 -- I 34014 Trieste, Italy
\vskip 0.5 truecm

{\bf L. Salasnich}\\
Dipartimento di Fisica dell'Universit\`a di Firenze,\\
INFN, Sezione di Firenze,\\
Largo E. Fermi, I 50125 Firenze, Italy

\vskip 1. truecm
{\it Preprint DFPD/93/TH/47} \\
To be published in Nuovo Cimento B
\end{center}

\newpage

\begin{center}
{\bf Abstract}
\end{center}

\par
A test on the numerical accuracy of the semiclassical approximation as a
function of the principal quantum number has been performed for the
Pullen--Edmonds model, a two--dimensional, non--integrable, scaling
invariant perturbation of the resonant harmonic oscillator. A perturbative
interpretation is obtained of the recently observed phenomenon of the
accuracy decrease on the approximation of individual energy levels at the
increase of the principal quantum number. Moreover, the accuracy provided
by the semiclassical approximation formula is on the average the same as
that provided by quantum perturbation theory.

\newpage

\par
Recently, there has been considerable renewed interest in the various
aspects of the Semi--Classical Approximation (SCA), a powerful motivation
behind that being the problem of the so--called quantum chaos (see for
example references [1,2,3,4,5]). An important aspect is represented by the
quantum energy levels, and in this connection one recent work [6] shows
that the predictions of individual levels by SCA (by this we mean the
Bohr--Sommerfeld formula, or one of its generalizations to the non--integrable
case, such as EBK; see e.g. [3,4]), worsen as the quantum number increases,
contrary to the naive expectation. We argue that this result can be
interpreted as follows: if $\hbar$, no matter how small, is kept fixed, the
SCA on the individual levels has the meaning of a perturbation theory (PT)
in $\hbar$. Therefore the accuracy of the approximation decreases for
higher levels (to get good agreement it is necessary, as is well known, to
implement the classical limit $\hbar \to 0$, $n \to \infty$, $n\hbar =I$
classical action; see e.g. [7]).
\par
The aim of this paper is to clarify this point, from the theoretical
standpoint and from the computational one as well, considering a scaling
invariant potential, which makes ordinary quantum PT strictly equivalent to
a power expansion in $\hbar$. We do actually observe that, for $\hbar$
fixed, the perturbation strength has to be decreased to keep the accuracy
at a constant value as the quantum number increases; however we also
observe that the algorithm provided by the appropriate SCA is always
comparable to the algorithm provided by ordinary quantum PT. A very good
agreement between the "exact" eigenvalues, obtained by numerical
diagonalization of the Schroedinger operator, and the semi--classical ones,
is indeed observed in presence of high unperturbed degeneracy.
\par
The most significant examples to carry out this comparison are represented
by non--separable two dimensional systems exhibiting both regular and
irregular spectrum [8], i.e. in particular, non--uniform behaviour of the
level spacing, and among these the simplest one is the Pullen--Edmonds
model [9]. Its quantum hamiltonian is:
\beq
H=-{\hbar^2\over 2}({\partial^2 \over \partial q_{1}^2}-
{\partial^2 \over \partial q_{2}^2})+{1\over 2}(q_{1}^2+q_{2}^2)+
\chi q_1^2 q_2^2,
\eeq
where $m$ has been put equal to $1$. For $\chi =0$ (1) reduces to a
resonant two--dimensional harmonic oscillator of levels
$(n_1+n_2+1)\hbar=m_1\hbar$, $m_1=1,2,...$, of multiplicity $m_1$.
\par
The scaling transformation $q_1\to \sqrt{\hbar}q_1$,
$q_2\to\sqrt{\hbar}q_2$ yields the unitary equivalent operator:
\beq
\tilde{H}=-\hbar [{1\over 2} ({\partial^2 \over \partial q_{1}^2}-
{\partial^2 \over \partial q_{2}^2} )+{1\over 2}(q_{1}^2+q_{2}^2)+
\chi \hbar q_1^2 q_2^2 ].
\eeq
The coupling constant has become $\chi \hbar $, which clearly shows
equivalence between expansions in powers of $\chi$ or of $\hbar$ (an
analogous result holds for any other polynomial perturbation). Moreover,
the symmetry of the potential enables us to split the hamiltonian matrix,
computed on the harmonic oscillator basis, into submatrices reducing the
computer storage required. The matrix elements of (1) can be written:
\beqa
&&<n_{1}^{'}n_{2}^{'}|H|n_{1}n_{2}>=\hbar (n_{1}+n_{2}+1)
\delta_{n_{1}^{'}n_{1}}
\delta_{n_{2}n_{2}}+\chi {\hbar^2 \over 4}[\sqrt{n_{1}(n_{1}-1)}
\delta_{n^{'}_{1}n_{1}-2}+
\nonumber \\
&&+\sqrt{(n_{1}+1)(n_{1}+2)}\delta_{n^{'}_{1}n_{1}+2}+
(2n_{1}+1)\delta_{n^{'}_{1}n_{1}}]\times
\\
&&\times [\sqrt{n_2(n_2-1)}\delta_{n'_2 n_2-2}+
\sqrt{(n_2+1)(n_2+2)}\delta_{n'_2 n_2+2}+
(2n_2+1)\delta_{n'_2 n_2}]
\nonumber
\eeqa
and each submatrix can be labelled by the parity of the occupation numbers
$n_1$, $n_2$. We restrict from now on to the invariant subspace spanned by
$m_1$ even, i.e. $n_1$ and $n_2$ of opposite parity. The eigenvalues of $H$
in this subspace have constant multiplicity 2 [9]. Therefore the level
$m_1\hbar=2s\hbar$ splits into $s$ levels for $\chi >0$.
\par
The appropriate SCA is here provided by the Bohr--Sommerfeld quantization
of the resonant (or secular) canonical perturbation theory [10],
also known, in this particular case, as the Birkoff--Gustafson normal form
[11,12], which we now construct at first order. Starting from the classical
Pullen--Edmonds hamiltonian:
\beq
H_{cl}={1\over 2}(p_1^2+p_2^2)+{1\over 2}(q_1^2+q_2^2)+
       \chi q_1^2 q_2^2,
\eeq
we introduce the standard action--angle variables $(I,\theta )$ by
the canonical transformation:
\beq
\left\{
\ba{ccc}
q_{i}&=&\sqrt{2I_{i}}\cos{\theta_{i}} \\
p_{i}&=&\sqrt{2I_{i}}\sin{\theta_{i}}.
\ea
\right.
\;\;\; i=1,2.
\eeq
Then (4) becomes:
\beq
H_{cl}=I_{1}+I_{2}+4\chi I_{1}^{2}I_{2}^{2}
       \cos^{2}{\theta_{1}}\cos^{2}{\theta_{2}}.
\eeq
The second canonical transformation into the well known "slow" and "fast"
variables:
\beq
\left\{
\ba{ccc}
A_{1}&=&I_{1}+I_{2} \\
A_{2}&=&I_{1}-I_{2}
\ea
\right.
\;\;\;\;\;
\left\{
\ba{ccc}
\theta_{1}&=&\phi_{1}+\phi_{2} \\
\theta_{2}&=&\phi_{1}-\phi_{2},
\ea
\right.
\eeq
eliminates the dependence on the "slow action" $A_2$ in the unperturbed
part, so that the hamiltonian becomes:
\beq
H_{cl}=A_{1}+\chi (A_{1}^{2}-A_{2}^{2})
       \cos^{2}{(\phi_{1}+\phi_{2})}\cos^{2}{(\phi_{1}-\phi_{2})}.
\eeq
We now eliminate the dependence on the angles up to order $\chi^2$ by
resonant (or secular) canonical perturbation theory [10]. To eliminate the
dependence on the "fast angle" $\phi_{1}$ it is enough to average the
perturbation on this variable. This yields:
\beq {1\over
2\pi}\int_{0}^{2\pi}d\phi_{1}
\cos^{2}{(\phi_{1}+\phi_{2})}\cos^{2}{(\phi_{1}-\phi_{2})} ={1\over
8}(2+\cos{4\phi_2}),
\eeq
and thus:
\beq
{\bar H}_{cl}=A_{1}+{\chi \over 8}(A_{1}^{2}-A_{2}^{2})
(2+\cos{4\phi_2}).
\eeq
The dependence of $\phi_2$ on the perturbation part can now
eliminated by a further canonical transformation.
The Hamilton--Jacobi equation for the perturbation part
is in fact:
\beq
[A_{1}^{2}-({\partial S\over \partial
\phi_{2}})^{2}] (2+\cos{4\phi_{2}})=K,
\eeq
\beq
{\partial S\over \partial \phi_{2}}=\pm
\sqrt{ A_{1}^{2}(2+\cos{4\phi_{2}})-K\over 2+\cos{4\phi_{2}} },
\eeq
and thus the Hamiltonian (9) becomes:
\beq
{\bar H}_{cl}=B_{1}+{\chi \over 8}K(B_{1},B_{2}),
\eeq
where:
\beq
B_{1}=A_{1}, \;\;\;\;
B_{2}={1\over 2\pi}\oint d\phi_{2} {\partial S\over \partial \phi_{2}}.
\eeq
It appears from the structure of equation (12) that the motions generated by
the perturbation part of our system have the following qualitative
behaviour:
\beq
\ba{cc}
0<K<B_{1}^{2} \;\;\;\; & rotational \; motion
\\
K=B_{1}^{2}   \;\;\;\; & separatrix
\\
B_{1}^{2}<K<3B_{1}^{2} \;\;\;\; & librational \; motion.
\ea
\eeq
The appearance of a separatrix (which is not immediately obvious in the
$(p,q)$ coordinates) accounts as is well known (see e.g. [3]) for the
stochastic layers originating near it. This corresponds to local irregular
behaviour of the quantum spectrum; one of its manifestations is (see D.
Delande in [4]) the local shrinking of the level spacing and the tendency
to avoided crossing. The shrinking of the level spacing is best accounted
by the SCA, as we will discuss below.
\par
On the separatrix we have:
\beq
B_{1}^{2}(2+\cos{4\phi_{2}})=K,
\eeq
while in general:
\beq
B_{2}=\pm {2\over \pi}\int_{a}^{b}dx
\sqrt{ B_{1}^{2}(2+\cos{4x})-K\over 2+\cos{4x} },
\eeq
where:
\beq
\ba{cc}
a=0, \;\; b={\pi\over 2} \;\;\;\; & rotational \; motion
\\
a=\phi_{-}(K,B_{1}), \;\; b=\phi_{+}(K,B_{1})
\;\;\;\; & librational \; motion
\ea
\eeq
with:
\beq
\phi_{\pm}(K,B_{1})=\pm {1\over 4}\arccos ({K\over B_1^2}-2).
\eeq
Now the approximate hamiltonian (13) depends only on the
actions so that a semiclassical quantization formula for the
($m_1$ even part) of spectrum of the operator (1) can be obtained by a
straightforward application of the Bohr--Sommerfeld
quantization rules [10]. Set therefore:
\beq
\left\{
\ba{ccc}
I_{1}&=&(n_1+1/2)\hbar \\
I_{2}&=&(n_2+1/2)\hbar,
\ea
\right.
\eeq
whence, from (6):
\beq
\left\{
\ba{ccc}
A_{1}&=&(n_1+n_2+1)\hbar \\
A_{2}&=&(n_1-n_2)\hbar .
\ea
\right.
\eeq
Set:
\beq
\left\{
\ba{ccc}
A_{1}&=&m_1\hbar \\
A_{2}&=&m_2\hbar,
\ea
\right.
\eeq
by comparison of (19) and (20) we obtain:
\beq
\left\{
\ba{ccc}
m_{1}&=&n_1+n_2+1 \\
m_{2}&=&n_1-n_2,
\ea
\right.
\eeq
where $m_1=2,4,...$ and $m_2= \pm (m_1-1), \pm (m_1-3), \pm (m_1-5),...$ .
\par
Finally:
\beq
B_{1}=m_{1}\hbar , \;\; B_{2}=m_{2}\hbar ;
\eeq
then the semiclassical approximation to the quantum spectrum is:
\beq
E_{m_{1},m_{2}}=m_{1}\hbar +{\chi \over 8}K(m_{1}\hbar , m_{2}\hbar ),
\eeq
where $K$ is implicitly defined by the relation:
\beq
m_{2}\hbar =\pm {2\over \pi}\int_{a}^{b}dx
\sqrt{ (m_{1}\hbar )^{2}(2+\cos{4x})-K\over 2+\cos{4x} },
\eeq
and:
\beq
\ba{cc}
a=0, \;\; b={\pi\over 2} \;\;\;\; & 0<K<(m_{1}\hbar )^{2}
\\
a=\phi_{-}(K,B_{1}), \;\; b=\phi_{+}(K,B_{1})
\;\;\;\; & (m_{1}\hbar )^{2}<K<3(m_{1}\hbar )^{2}.
\ea
\eeq
\par
Remark that for $|m_2|<[\alpha m_1]$ we obtain the quantization of the
rotational motions, while for $|m_2|>[\alpha m_1]$ ($[x]$=integer part of
$x$) we have the quantization of the librational ones. Here, by (17):
\beq
\alpha ={2\over \pi}\int_0^{\pi\over 2}dx \sqrt{1+\cos{4x}\over 2
+\cos{4x}}\simeq 0.602.
\eeq
\par
Moreover, it immediate to see that for $m_1$ fixed the function $K$, and
hence the semiclassical energy $E_{m_1,m_2}$, is a decreasing function of
the secondary quantum number $m_2$. It is furthermore proved in [13] that
(25) coincides with the exact quantum spectrum up to terms of order $\hbar$
and $\chi^2$. The numerical computations (see Fig. 4 below) show that at
order 1 in $\chi$ the corrections of order $\hbar$ affect at most the eight
decimal figure.
\par
The "exact" levels have been computed, and compared with the semiclassical
ones as well as with the levels computed by degenerate first order quantum
perturbation theory [14], for $m_1=1,...,60$ at $\hbar =0.1$ and for
different values of $\chi$ (given the degeneracy, this is equivalent to
compute $1800$ different levels). The results obtained for $m_1=30$,
$\hbar=0.1$, $\chi =10^{-3}$ and $m_1=60$, $\hbar =0.1$, $\chi =10^{-5}$
are shown in Figure 1 and Figure 2, respectively. The local shrinking of
the spacing, reproduced by both methods, can be immediately noticed; remark
that the corresponding semiclassical levels are those near the separatrix
(by (28), $m_2 \sim 18$ and $m_2 \sim 36$, respectively).
\par
In Figure 3 the function:
\beq
\Delta =|E^{Ex}-E^{Sc}|
\eeq
{\it vs} $m_1$ is plotted for $m_1\chi =1$; this shows that, if the
coupling constant is decreased in inverse proportion to the principal
quantum number the accuracy of SCA not only remains constant but actually
improves, as anticipated because the scaling invariance makes the limit
$m_1\to \infty$, $\chi \to 0$, $m_1\chi \to const$ equivalent to the
classical limit $m_1\to \infty$, $\hbar \to 0$, $m_1\hbar \to const$.
\par
In Figure 4 the accuracies obtained thorough semiclassical and quantum
first order perturbation theories are compared for $m_1=60$, $\chi =10^{-
5}$, $\hbar =0.1$, and in Fig. 5 the difference between the two
perturbation theories is plotted (remark that the energy decreases as $m_2$
increases). As can be seen the agreement with the "exact" levels is very
good and the accuracy is on the average the same. Remark however that, as
it should be expected (the Bohr--Sommerfeld rules take no account of
tunneling [15]), the lowest accuracy of the SCA is reached near $m_2=36$
which corresponds to the levels near the separatrix: for those levels the
quantum PT is therefore better than SCA.

\begin{center}
*****
\end{center}
\par
The authors are greatly indebted to Dr. Stefano Isola for many useful
discussions and to Mr. G. Salmaso for his valuable computational
assistance.

\newpage

\parindent=0.pt
\section*{Figure Captions}
\vspace{0.6 cm}

Figure 1: Comparison between the "exact" levels (a), the semi--classical
ones (b), and the levels obtained by first order perturbation theory (c),
for $\chi =10^{-3}$, $\hbar =0.1$, $m_1=30$.

Figure 2: Comparison between the "exact" levels (a), the semi--classical
ones (b), and the levels obtained by first order perturbation theory (c),
for $m_1=60$, $\chi =10^{-5}$.

Figure 3: The difference $\Delta$ between the "exact" levels and the
semi--classical ones {\it vs} $m_1$, with $m_2=m_1-1$ and $m_1\hbar =1$.

Figure 4: (a) The difference between the "exact" levels and the
semi--classical ones; (b) the difference between the "exact" levels and the
first order quantum PT ones; ($m_1=60$, $\hbar =0.1$, $\chi =10^{-5}$).

Figure 5: The difference between the
semi--classical levels and the first order quantum PT ones;
($m_1=60$, $\hbar =0.1$, $\chi =10^{-5}$).

\newpage

\section*{References}
\vspace{0.6 cm}

[1] V.P. Maslov and M.V. Fedoriuk: {\it Semi--Classical Approximation in
Quantum Mechanics} (Reidel Publishing Company, 1981)

[2] A.B. Migdal: {\it Qualitative Methods in Quantum Theory} (Benjamin,
1977)

[3] M.C. Gutzwiller: {\it Chaos in Classical and Quantum Mechanics}
(Springer--Verlag, 1991);
A.R. Rau: Rev. Mod. Phys. {\bf 64}, 623 (1992)

[4] D. Delande: {\it Chaos and Quantum Physics}, in Les
Houches Summer School 1989, Ed. M.J. Giannoni, A. Voros and J. Zinn--Justin
(Elsevier Science Publishing, 1989)

[5] P.A. Braun: Rev. Mod. Phys. {\bf 65}, 115 (1993)

[6] T. Prosen and M. Robnik: J. Phys. A {\bf 26}, L37 (1993)

[7] G. Alvarez, S. Graffi and H.J. Silverstone: Phys. Rev. A {\bf 38}, 1697
(1988)

[8] I. C. Percival: J. Phys. B {\bf 6}, L229 (1973)

[9] R.A. Pullen, R.A. Edmonds: J. Phys. A {\bf 14}, L477
(1981)

[10] M. Born: {\it Mechanics of the Atom} (Bell, 1960)

[11] W.P. Reinhardt: in {\it The Mathematical Analysis
of Physical Systems}, Ed. R.E. Mickens and R. Van Nostrand (1984)

[12] T. Uzer, D.W. Noid ans R.A. Marcus: J. Chem. Phys. {\bf 79}, 4412
(1983)

[13] S. Graffi: in {\it Probabilistic Methods in Mathematical Physics}, Ed.
F. Guerra, M.I. Loffredo and C. Marchioro (World Scientific, 1992)

[14] L. Landau and E. Lifshitz: {\it Quantum Mechanics: Non Relativistic
Theory} (1978)

[15] A.M. Ozorio de Almeida: J. Phys. Chem. {\bf 88}, 6139 (1984)

\end{document}